\documentclass{jps-cp}
\usepackage{txfonts} 
\usepackage{subfigure}%
\usepackage{url}
\usepackage{color}

\title{Characterization of SiPM Optical Crosstalk and its Dependence on the Protection-Window Thickness}

\author{Yuki~\textsc{Nakamura}, Akira~\textsc{Okumura}, Hiroyasu~\textsc{Tajima}, Nobuhito~\textsc{Yamane}, and Anatolii~\textsc{Zenin}}

\inst{Institute for Space--Earth Environmental Research, Nagoya University, Furo-cho, Chikusa-ku, Nagoya 464-8601, Japan}

\email{nakamura.yuki@isee.nagoya-u.ac.jp (Y.~N.), oxon@mac.com~(A.~O.)}

\recdate{April 17, 2019}

\abst{Owing to their high photon detection efficiency, compactness, and low operating voltage, silicon photomultipliers (SiPMs) have found widespread application in many fields, including medical imaging, particle physics, and high-energy astrophysics. However, the so-called optical crosstalk (OCT) phenomenon of SiPMs is a major drawback to their adoption. Secondary infrared photons are emitted inside the silicon substrate spontaneously after the avalanche process caused by the primary incident photons, and they can be detected by the surrounding photodiodes. As a result large output pulses that are equivalent to multiple photoelectrons are observed with a certain probability (OCT rate), even for single-photon events, making the charge resolution worse and increasing the rate of accidental triggers by single-photon events in applications such as atmospheric Cherenkov telescopes. In our previous study, we found that the OCT rates of single-channel SiPMs was dependent on the thickness of their protection resin window, which may be explained by photon propagation inside the resin. In the present study, we measured the OCT rate of a multichannel SiPM and those of neighboring channels caused by photon propagation. Both OCT rates were found to be dependent on the protection-window thickness. We report our OCT measurements of a multichannel SiPM and comparisons with a ray-tracing simulation.}

\kword{Silicon photomultipliers (SiPMs), ray-tracing simulation, Cherenkov telescopes}

\begin{document}
\maketitle

\section{Introduction}
Silicon photomultipliers (SiPMs) are a gathering of multiple avalanche photodiode cells that operate in a Geiger mode (G-APD). While a single G-APD cell produces a constant charge signal that does not depend on the number of photons hitting the cell, SiPMs are capable of counting the number of incident photons by arranging multiple G-APD cells in parallel. Nowadays SiPMs are used in a variety of applications, such as in medical imaging, particle physics, and high-energy astrophysics owing to their robustness, low cost per channel, low operating voltage, and high tolerance to bright illumination. However, SiPMs have a detrimental phenomenon, the so-called optical crosstalk (OCT), that can negatively affect the system performance in some applications.

OCT is a phenomenon whereby the avalanche process in a single G-APD cell emits secondary infrared photons that are then detected by the surrounding cells with a certain probability (OCT rate). As a result, the output chage from the SiPM channel is not constant, even for a single-photoelectron (1-p.e.) event, and it is proportional to one plus the number of detected secondary photons. For very-high-energy ground-based gamma-ray telescopes (also referred to as Cherenkov telescopes), which detect UV--visible photons from Cherenkov radiation in atmospheric showers, OCT can degrade the energy resolution of the primary particles, because the number of Cherenkov photons is approximately proportional to the energy. Furthermore, OCT can increase the rate of accidental triggers, which are caused by photons randomly coming from the night sky. This is because a single photon may generate a large output pulse equivalent to multiple photons, which cannot be distinguished from temporally correlated Cherenkov photons. Thus the level-1 trigger threshold needs to be set higher than preferred, resulting in a higher energy threshold and lower detection sensitivity to celestial gamma-ray sources.

The use of SiPMs as the focal-plane camera pixels of Cherenkov telescopes was first realized by the FACT telescope\cite{cite:FACT}. As of 2019, several SiPM camera prototypes are also being developed for the Cherenkov Telescope Array (CTA)\cite{cite:cta} with the aim to build compact and less expensive cameras with thousands of pixels that can have a high tolerance to bright moon nights.

\begin{figure}[tbp]
  \centering
  \includegraphics[clip, width=8cm]{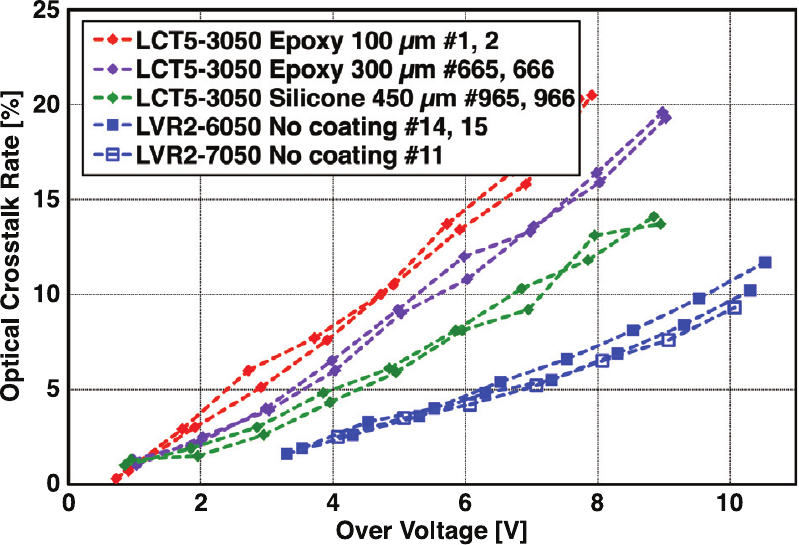}
  \caption{The OCT rates of five different SiPM products as functions of over voltage (defined to be the supplied bias voltage minus the breakdown voltage). The thicker the protection window, the lower OCT rate; but, those with no protection coating exhibit the lowest OCT rate. The figure is taken from \cite{cite:sigle-pix-oct} (reproduced under CC BY 4.0, see reference for details).}
  \label{fig:single-pix-oct}
\end{figure}

\begin{figure}[tbp]
  \centering
  \includegraphics[width=12cm]{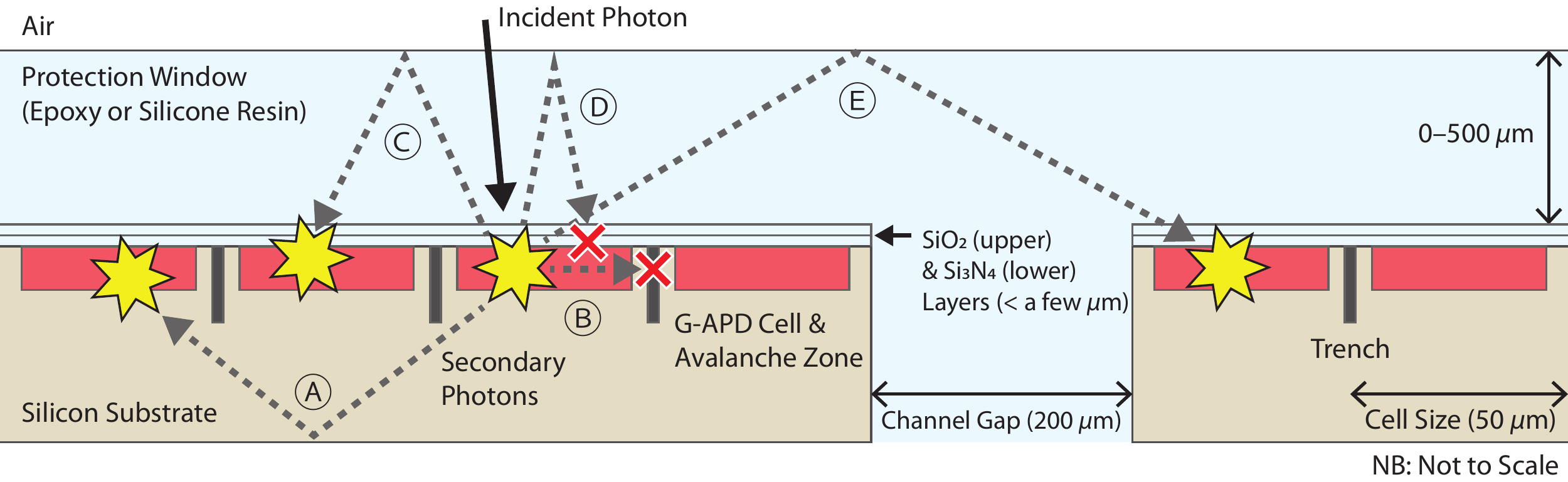}
  \caption{A schematic diagram of possible OCT photon paths inside a multichannel SiPM. After the avalanche process is triggered by the first incident photon (solid-line arrow), secondary photons may propagate and be detected by surrounding cells via several paths. Path A is due to Fresnel reflection at the boundary between the silicon substrate and the air. Path B is a direct path to an adjacent G-APD cell but mostly blocked by a trench. Path C is by Fresnel reflection at the boundary between the protection window and the air. Path D is similar to Path C, but the photon is reflected back to the first G-APD cell, which is already saturated. Path E is also similar to Path C, but it may be detected by neighboring channels in the case of multichannel SiPMs.}
  \label{fig:OCT-paths}
\end{figure}

In our previous study, in which an extensive characterization of various SiPM products was conducted for the CTA, we found that the OCT rate of single-channel SiPMs had a coating-thickness dependence, as shown in Fig.~\ref{fig:single-pix-oct}\cite{cite:sigle-pix-oct}. This dependence may be explained by photon propagation in the protection window, as discussed in the paper. However, a firm conclusion could not be derived because the OCT rate comparison was made for different products and resin materials, and because escaping photons (illustrated as path E in Fig.~\ref{fig:OCT-paths}) could not be confirmed with single-channel SiPMs.

Fig.~\ref{fig:OCT-paths} illustrates the possible paths of secondary photons propagating in the protection window and in the silicon substrate. It supports our photon propagation scenario and explains Fig.~\ref{fig:single-pix-oct} as follows.

Secondary photons that directly enter the neighboring cells (path B) can be blocked by the so-called trench structure developed in recent SiPM products \cite{cite:trench}. When the protection window is very thin or does not exist, the number of photons propagating in path D increases (i.e., path C or E does not occur), and the OCT rate decreases because the first G-APD cell is already saturated. By increasing the protection window thickness, we can make the number of photons propagating in path E larger than that in path C, and thus the OCT rate of a single-channel SiPM is reduced. However, if this scenario is correct, path E may generate OCT pulses in neighboring channels for multichannel SiPMs that are planned to be used in CTA cameras.

To characterize and understand the OCT rate of multichannel SiPMs and to confirm the possible photon paths in Fig.~\ref{fig:OCT-paths}, we measured the OCT rate of  multichannel SiPM S13361-6805 ($8\times8$ channels, $3\times3\ \mathrm{mm}^2$ each, 50-$\mu$m cells, no protection coating, Hamamatsu Photonics). To eliminate the possible systematic uncertainty due to different resin materials, we used thin glass plates and optical grease to emulate the different protection-window thickness.

In Section~\ref{sec:single}, we show the OCT rate for different (glass) window thickness to confirm the results of the previous study. In Section~\ref{sec:neighboring}, we present OCT events occurring in neighboring channels and their rates for the first time. In Section~\ref{sec:ray-tracing}, we also compare our measurements with a simple ray-tracing simulation and provide a conclusion in Section~\ref{sec:conclusion}.

\section{Optical Crosstalk Rate of a Single SiPM Channel}
\label{sec:single}

\begin{figure}[tbp]
  \centering
  \includegraphics[width=10cm]{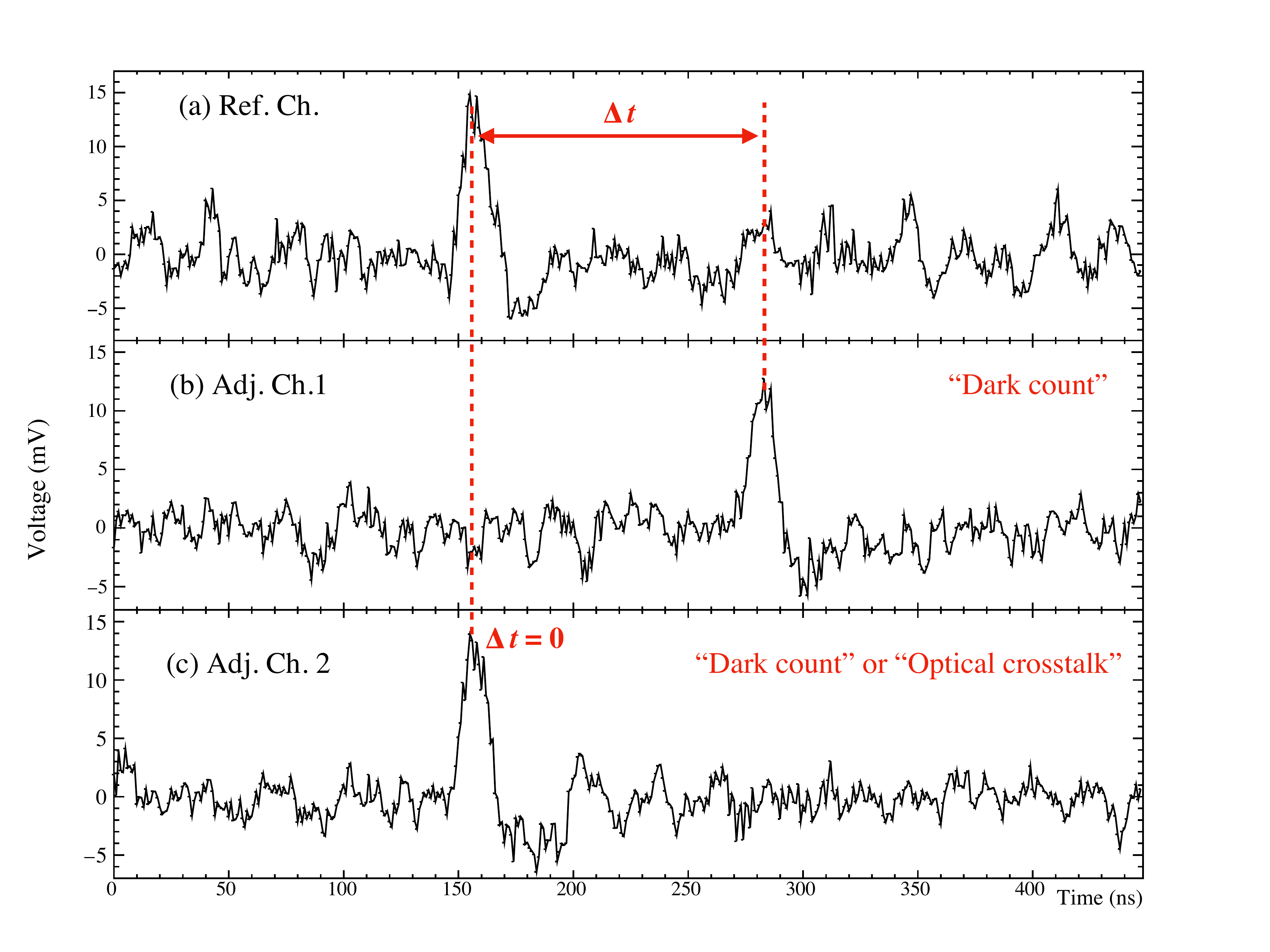}
  \caption{Example SiPM waveforms recorded with TARGET~7 chips: (a) 1-p.e. pulse in the reference channel is seen at around $155$~ns. (b) Same as (a) but for Adjacent Channel 1. The counts in the reference channel and this channel are not coincident. (c) Same as (a) but for Adjacent Channel 2. This may be temporally correlated with the pulse in the reference channel.}
  \label{fig:waveforms}
\end{figure}

The OCT rate of a particular single channel (hereafter referred to as the reference channel, see Fig.~\ref{fig:neighbor-oct-2D}) can be measured by taking the charge distribution of the channel output in a certain time window. Deriving the ratio between the numbers of 1-p.e. and 2-p.e. events under an assumption that the 2-p.e. events do not follow a pure Poisson distribution, the excess number of 2-p.e. events is considered to be due to OCT. Output waveforms were recorded with a 1 GSa/s sampling circuit named TARGET~7, a member of TARGET family developed for CTA \cite{cite:TARGET, cite:TARGET5}. A single TARGET~7 chip is able to record 16 input channels. To record all the 64 output channels simultaneously, a TARGET~7 camera module, equipped with four TARGET~7 chips, was used for this study. Example waveforms can be seen in Fig~\ref{fig:waveforms}.

\begin{figure}[tbp]
  \centering
  \includegraphics[width=8cm]{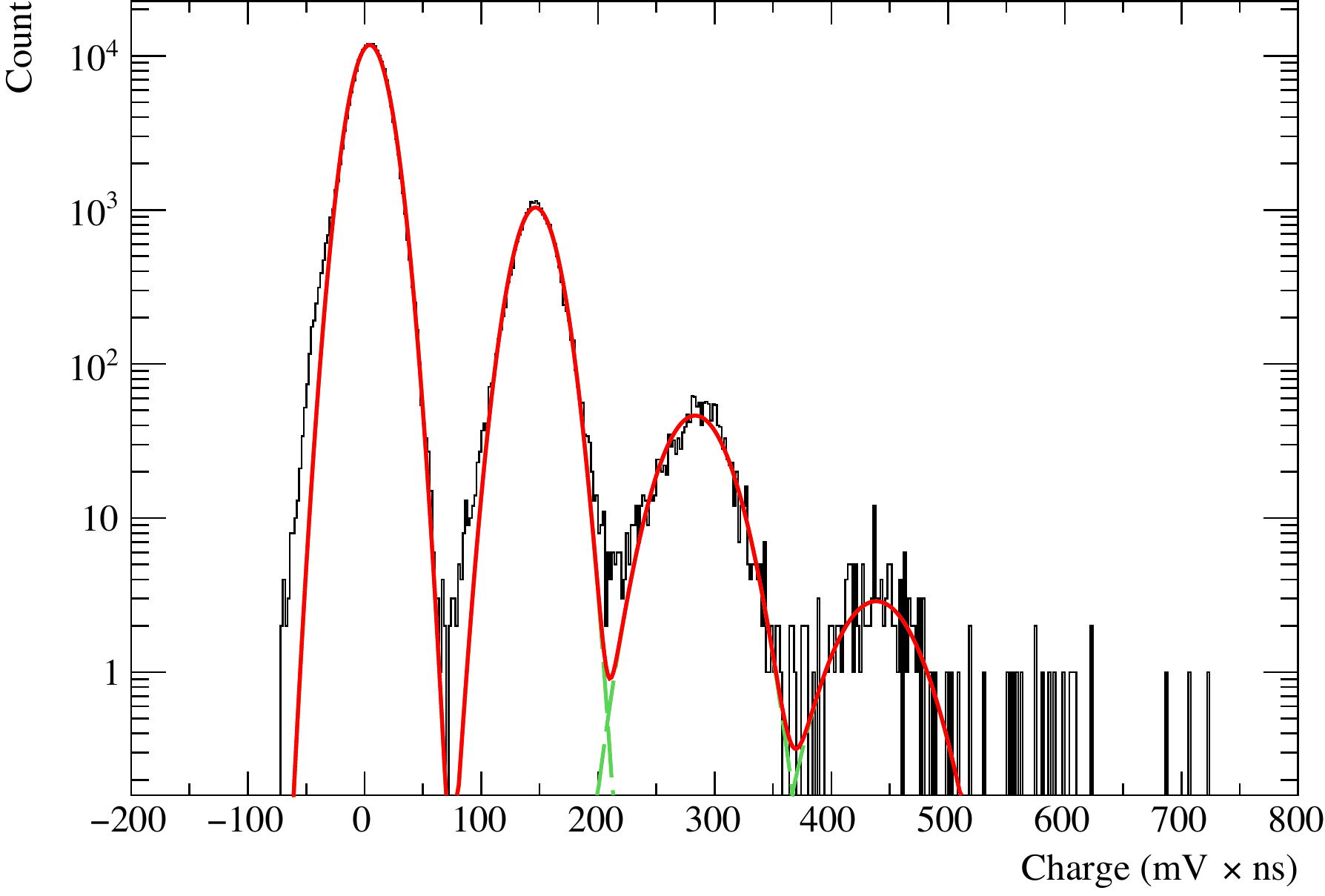}
  \caption{The charge distribution of the reference channel (solid black line), the best-fit multiGaussian (dashed green lines), and their sum (solid red line). Five obvious peaks can be seen from left to right: pedestal, 1~p.e., 2~p.e., and 3~p.e.. Systematic uncertainty in calculating the OCT rate due to the non-Gaussianity in this distribution is shown with vertical brackets in Fig.~\ref{fig:thickness-dep}.}
  \label{fig:phd}
\end{figure}

Fig.~\ref{fig:phd} shows a charge distribution of the reference channel, to which multiGaussian are fitted to derive the OCT rate,
\begin{equation}
R_\mathrm{OCT} = 1 - \frac{1 - \frac{N(\geq2~\mathrm{p.e.})}{N(\geq1~\mathrm{p.e.})}}{\exp{(-f_{\mathrm{DCR}} \Delta t_{\mathrm{ps}})}},
\end{equation}
where $N(\geq2~\mathrm{p.e.})$ and $N(\geq1~\mathrm{p.e.})$ are the numbers of events with a charge equivalent to 2~p.e. or more and 1~p.e. or more, respectively, $f_{\mathrm{DCR}}$ is the dark count rate and $\Delta t_{\mathrm{ps}}$ is the maximum time window in which two accidental dark counts cannot be temporally resolved in our deconvolution analysis\cite{cite:deconvolution}.

\begin{figure}[tbp]
  \centering
  \subfigure[]{%
    \label{fig:thickness-dep}
    \includegraphics[width=8cm]{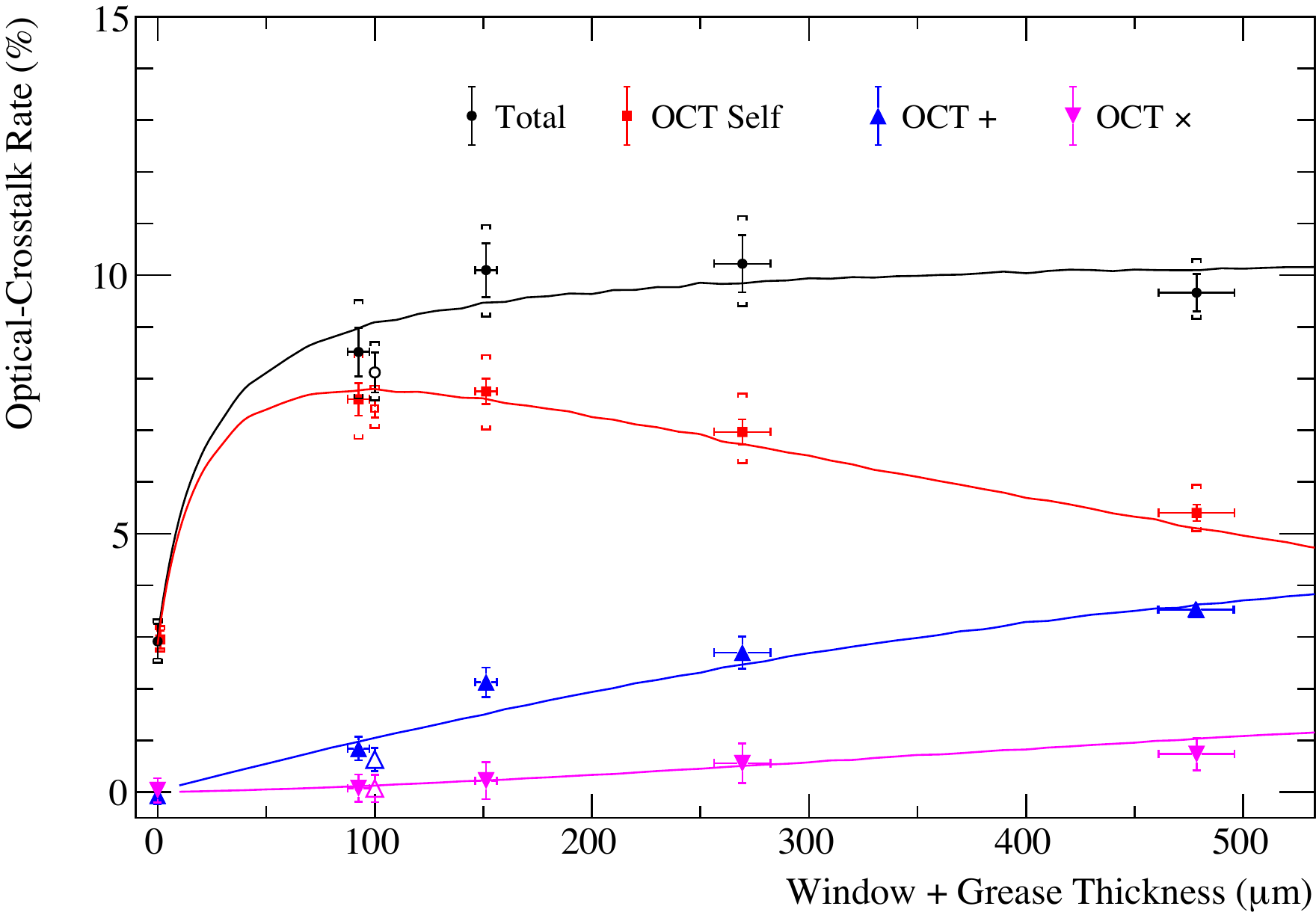}
  }
  \subfigure[]{%
    \label{fig:neighbor-oct-difinition}
    \includegraphics[width=4cm]{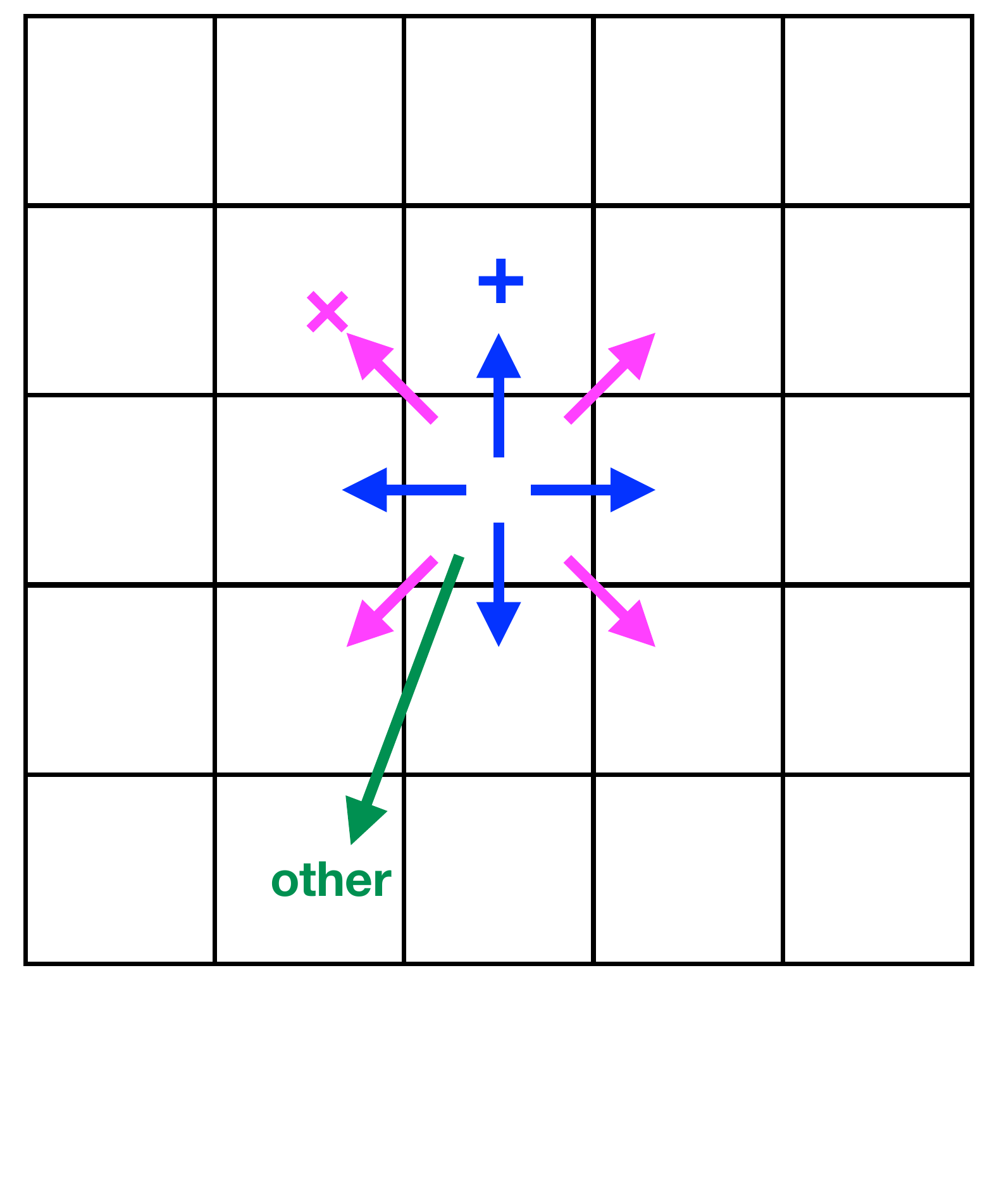}
  }
  \caption{(a) OCT rate vs. window thickness. The OCT rate in the reference channel (OCT self), that in neighboring channels at ``$+$'' and ``$\times$'' directions (OCT$_+$ and OCT$_\times$),  and the total. Ray-tracing simulations are shown with solid curves. Open symbols show the OCT rates of the same SiPM product with a 100-$\mu$m thick silicone resin to verify our window emulation method using thin glass plates. (b) Definition of ``$+$'' and ``$\times$'' directions.}
\end{figure}

The OCT rate of the reference channel with various dummy protection-window thickness is shown in Fig~\ref{fig:thickness-dep} (solid red squares). It is at the minimum and at the maximum when the thickness is $0$~$\mu$m (i.e., no glass plate is put) and $\sim100$~$\mu$m, respectively, and gradually decreases as the window becomes thicker. This confirms the similar trend found in the previous study (see Fig.~\ref{fig:single-pix-oct}).

\section{Optical Crosstalk to Neighboring Channels}
\label{sec:neighboring}

We measured the OCT rates in neighboring channels to test the OCT photon propagation pathways depicted in Fig.~\ref{fig:OCT-paths}. If OCT propagation to neighboring channels according to path E exists, coincident dark counts may be observed in the adjacent channels, as shown in Figs.~\ref{fig:waveforms}(a) and \ref{fig:waveforms}(c). Assuming that the first dark count happened in the reference channel, the 1-p.e. pulse in Adjacent Channel 2 is considered to be another dark count in the latter channel or an OCT photon detected right after the dark count in the reference channel. For individual pulses, we cannot determine which of the two scenarios is correct. On the contrary, for the pulse in Fig.~\ref{fig:waveforms}(b), it is not temporally correlated with the other two pulses and can be considered to be a dark count occurring in Adjacent Channel 1.

\begin{figure}[tbp]
  \centering
  \includegraphics[width=8cm]{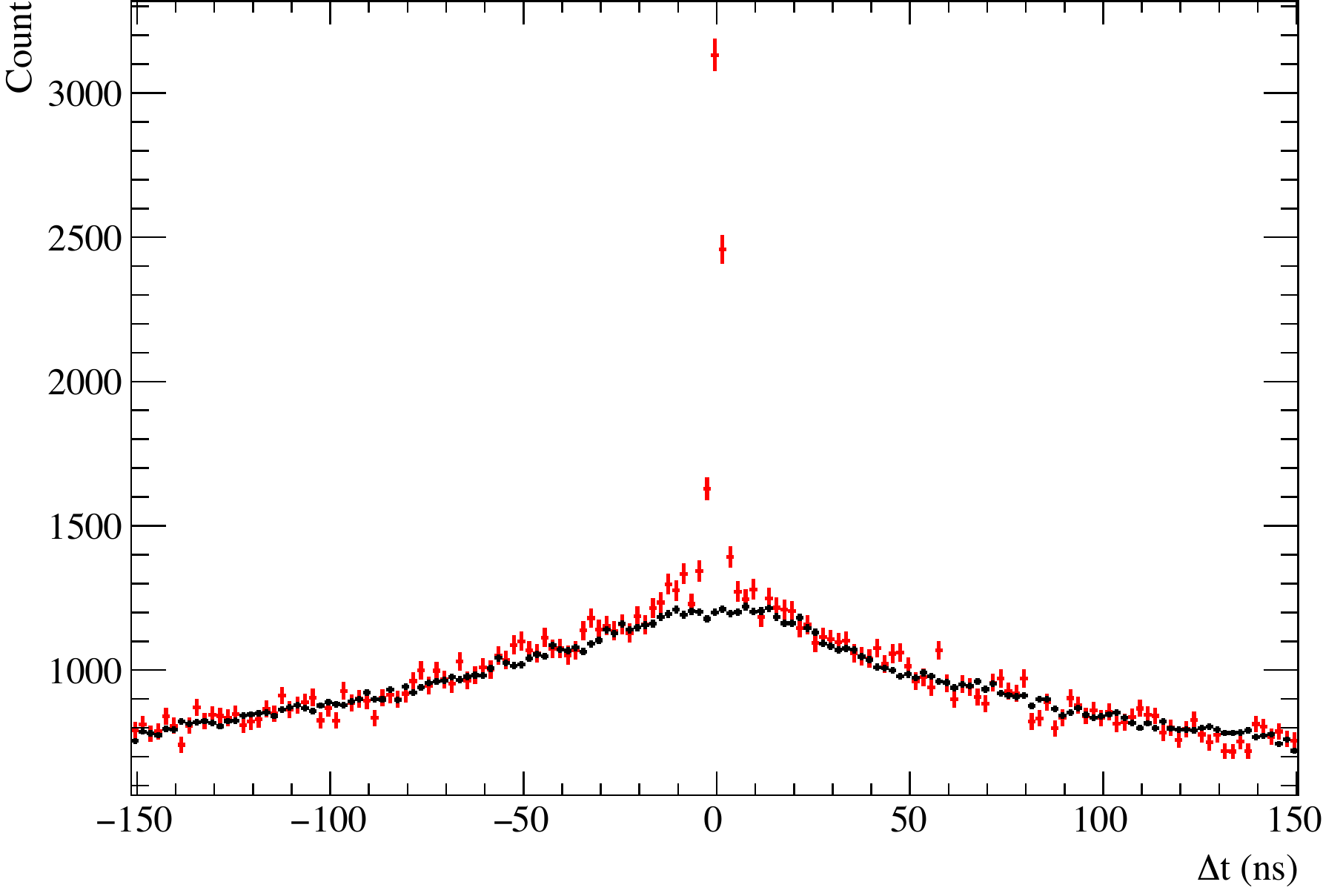}
  \caption{$\Delta t$ distributions of the reference channel and a neighboring channel. The red histogram shows $\Delta t$ for the two closest pulses appeared in the reference and neighboring channels. The black histogram was made from waveforms without any pulse in the reference channel ($t_0$ was randomly chosen in $\pm100$~ns).}
  \label{fig:dt-dist}
\end{figure}

The two possibilities can be later separated by making distributions of the time difference of photoelectron detection timings ($\Delta t$ ($\equiv t_0- t_1$) distribution, $t_0$ is the detection timing in the reference channel, $t_1$ is in a neighboring channel). Fig.~\ref{fig:dt-dist} shows two $\Delta t$ distributions: The red distribution was made from only waveforms in which a photoelectron was detected in a certain time window in the reference channel. The black distribution was made from waveforms without any pulse in the reference channel and $t_0$ was randomly chosen in $\pm100$~ns, resulting in an exponential-like distribution.

The excess around $\Delta t = 0$ in the red histogram was caused by OCT photons propagating to the neighboring channels. Integrating this excess within $\pm32$~ns (wider than the timing resolution in our peak search), the OCT rate in the neighboring channel, which originates from dark counts in the reference channel, can be calculated.

\begin{figure}[tbp]
  \centering
  \subfigure[]{%
    \label{fig:64-ch-MPPC}
    \includegraphics[width=4.5cm]{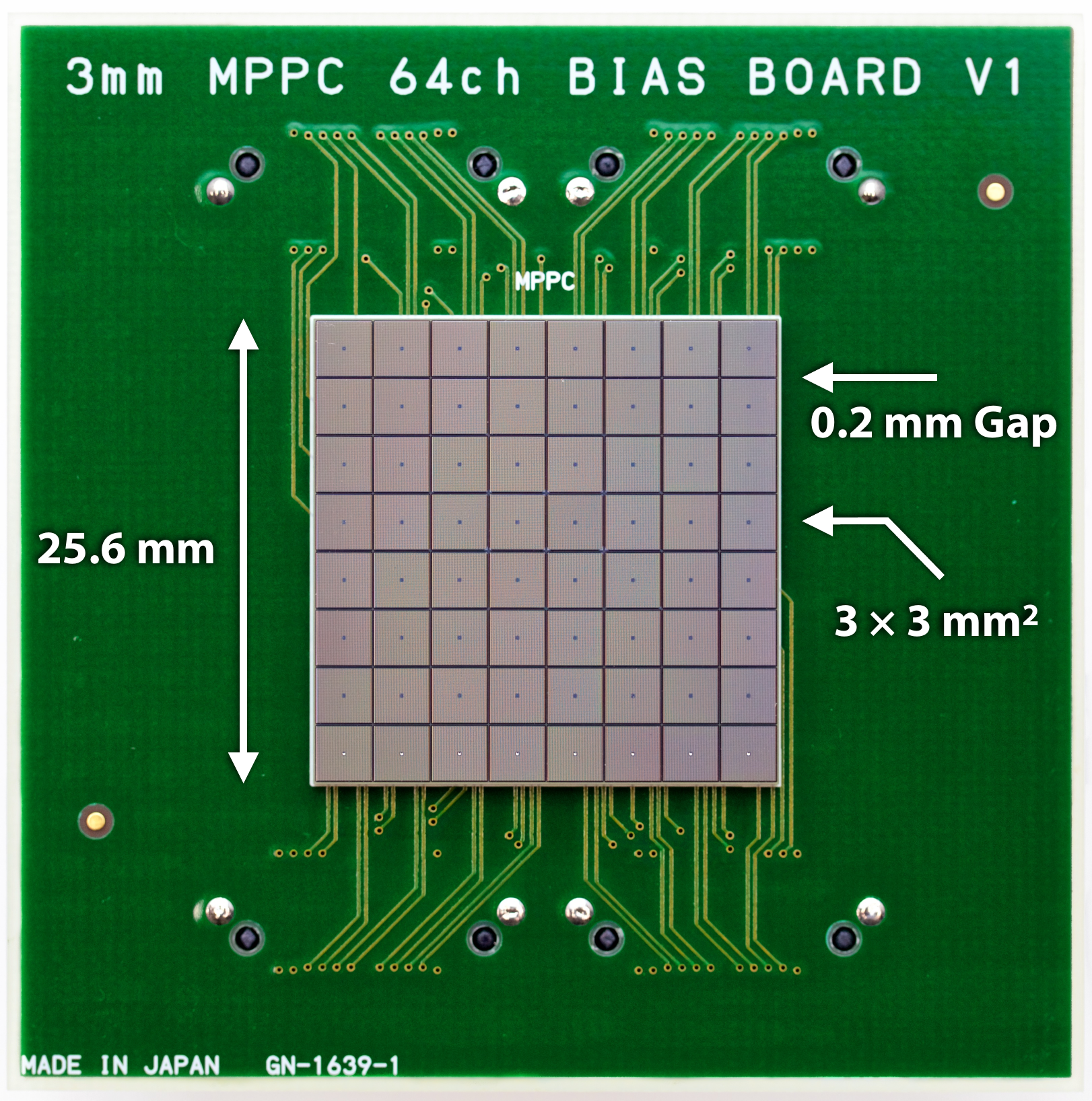}
  }
  \subfigure[]{%
    \label{fig:neighbor-oct-2D}
    \includegraphics[width=6.5cm]{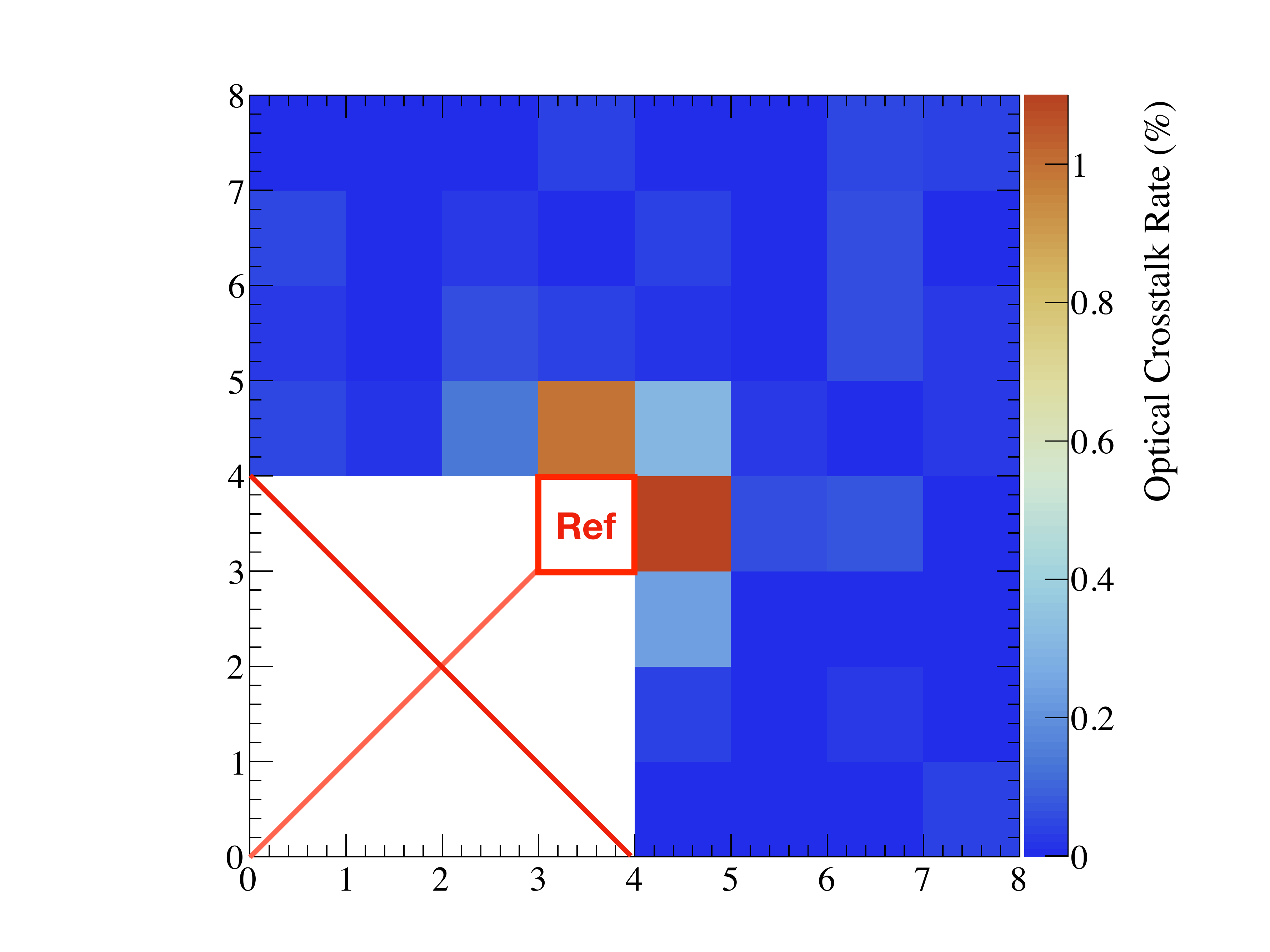}
  }

  \caption{(a) $8\times8$ channel SiPM array (Hamamatsu Photonics S13361-6805). (b) An example 2D distribution of neighboring-channel OCT rates. The blank 15 channels were excluded in the analysis to avoid common electrical noise of the TARGET ASIC connected to the reference channel.}
\end{figure}

Fig.~\ref{fig:neighbor-oct-2D} shows the $8\times8$ channel SiPM and a 2D distribution of neighboring-channel OCT rates originating from dark counts in the reference channel. Repeating measurements with different window thicknesses, we obtained the window-thickness dependence of the OCT rates (Fig~\ref{fig:thickness-dep}). Neighboring-channel OCT occurs in adjacent channels with non-negligible rates ($>\sim1$\%) when the window is $\sim100$~$\mu$m or thicker and the total OCT rate summed from the reference and adjacent channels is roughly constant in this thickness region. These findings confirmed that OCT photons indeed propagate to neighboring channels due to reflection at the media boundary between the window and the ambient air.

\section{Ray-tracing Simulation}
\label{sec:ray-tracing}
The measurement results shown in Fig.~\ref{fig:thickness-dep} and OCT photon paths shown in Fig.~\ref{fig:OCT-paths} were also verified by ray-tracing simulations. We used the ROBAST library\cite{cite:robast} to build the multichannel SiPM and window geometries, inside which the OCT photon propagation was simulated assuming a simple Fresnel reflection at the media boundaries and isotropic and uniform OCT photon emission from all the APD cells in the reference channel. As shown in Fig.~\ref{fig:thickness-dep}, the simulation results after rough normalization by eye are consistent with the measurements.

\section{Conclusion}
\label{sec:conclusion}
We measured the neighboring-channel OCT rates of a multichannel SiPM and its window-thickness dependence for the first time. The ``self OCT'' rate was the highest at around $100$~$\mu$m and it can be decreased by removing the window or by increasing the thickness of the window. The neighboring-channel OCT rates increased as the window thickness increased and neighboring-channel OCT could only be reduced by removing the window entirely. Thus, removing the protection window is the preferred way to significantly suppress the total OCT rate in multichannel SiPM applications like CTA unless the SiPM surface handling is problematic.


\begin{thebibliography}{9}
\bibitem{cite:trench} D.~McNally, V.~Golovin, \textit{Nucl. Instrum. Methods Phys. Res., Sect. A} \textbf{610}, 150--153 (2009).

\bibitem{cite:FACT} H. Anderhub, \textit{et~al.}, \textit{J. Instrum.} \textbf{8} P06008 (2013).

\bibitem{cite:cta} B.~S.~Acharya, \textit{et~al.}, \textit{Astropart. Phys.} \textbf{40}, 3--18, (2013).

\bibitem{cite:hamamatsu} Hamamatsu~Photonics. in \textit{Opto-Semiconductor handbook}, Chap.~3.\\
  \url{https://www.hamamatsu.com/resources/pdf/ssd/03_handbook.pdf}

\bibitem{cite:sigle-pix-oct} A.~Asano, \textit{et~al.}, \textit{Nucl. Instrum. Methods Phys. Res., Sect. A} \textbf{912}, 177--181 (2018).

\bibitem{cite:TARGET} K.~Bechtol, \textit{et~al.}, \textit{Astropart. Phys.} \textbf{36}, 156--165 (2012).
  
\bibitem{cite:TARGET5} A.~Albert, \textit{et~al.}, \textit{Astropart. Phys.} \textbf{92} 49--61 (2017).

\bibitem{cite:deconvolution} H.~Otono, \textit{et~al.},  \textit{Nucl. Instrum. Methods Phys. Res., Sect. A}, \textbf{610}, 397--399 (2009).

\bibitem{cite:robast} A.~Okumura, K.~Noda, and C.~Rulten, \textit{Astropart. Phys.} \textbf{76} 38--47 (2016).

\end{thebibliography}
\end{document}